\documentclass[]{spie}  

\usepackage{amsmath,amsfonts,amssymb}
\usepackage{graphicx}
\usepackage[colorlinks=true, allcolors=blue]{hyperref}
\usepackage{lineno}
\usepackage{setspace}
\title{ImageNomer: description of a functional connectivity and omics analysis tool and case study identifying a race confound}

\author[a]{Anton Orlichenko}
\author[b]{Grant Daly}
\author[a]{Ziyu Zhou}
\author[c]{Anqi Liu}
\author[c]{Hui Shen}
\author[c]{Hong-Wen Deng}
\author[a]{Yu-Ping Wang}
\affil[a]{Department of Biomedical Engineering, Tulane University, New Orleans, LA, USA}
\affil[b]{College of Medicine, University of South Alabama, Mobile, AL, USA}
\affil[c]{School of Medicine, Tulane University, New Orleans, LA, USA}

\authorinfo{Further author information: (Send correspondence to Anton Orlichenko)\\Anton Orlichenko: E-mail: aorlichenko@tulane.edu}

\begin{document} 
\maketitle

\begin{abstract}
Most packages for the analysis of fMRI-based functional connectivity (FC) and genomic data are used with a programming language interface, lacking an easy-to-navigate GUI frontend. This exacerbates two problems found in these types of data: demographic confounds and quality control in the face of high dimensionality of features. The reason is that it is too slow and cumbersome to use a programming interface to create all the necessary visualizations required to identify all correlations, confounding effects, or quality control problems in a dataset. FC in particular usually contains tens of thousands of features per subject, and can only be summarized and efficiently explored using visualizations. To remedy this situation, we have developed ImageNomer, a data visualization and analysis tool that allows inspection of both subject-level and cohort-level demographic, genomic, and imaging features. The software is Python-based, runs in a self-contained Docker image, and contains a browser-based GUI frontend. We demonstrate the usefulness of ImageNomer by identifying an unexpected race confound when predicting achievement scores in the Philadelphia Neurodevelopmental Cohort (PNC) dataset, which contains multitask fMRI and single nucleotide polymorphism (SNP) data of healthy adolescents. In the past, many studies have attempted to use FC to identify achievement-related features in fMRI. Using ImageNomer to visualize trends in achievement scores between races, we find a clear potential for confounding effects if race can be predicted using FC. Using correlation analysis in the ImageNomer software, we show that FCs correlated with Wide Range Achievement Test (WRAT) score are in fact more highly correlated with race. Investigating further, we find that whereas both FC and SNP (genomic) features can account for 10-15\% of WRAT score variation, this predictive ability disappears when controlling for race. We also use ImageNomer to investigate race-FC correlation in the Bipolar and Schizophrenia Network for Intermediate Phenotypes (BSNIP) dataset. In this work, we demonstrate the advantage of our ImageNomer GUI tool in data exploration and confound detection. Additionally, this work identifies race as a strong confound in FC data and casts doubt on the possibility of finding unbiased achievement-related features in fMRI and SNP data of healthy adolescents.
\end{abstract}


\keywords{fMRI, functional connectivity, software, achievement score, race confound, PNC dataset}

\section{INTRODUCTION}
\label{sec:intro}  

Functional magnetic resonance imaging (fMRI) uses the blood oxygen level-dependent (BOLD) signal to identify regions of increased neural activity. \cite{Belliveau1991FunctionalMO} Functional connectivity (FC) is an fMRI-derived measure that quantifies the synchronization between BOLD signal in different regions of the brain. \cite{doi:10.1073/pnas.0135058100} It and similar measures have been used to predict age,\cite{10002422} sex,\cite{Zhang2020GenderDA} intelligence,\cite{Qu2021EnsembleMR} and disease status. \cite{Du2018ClassificationAP}\cite{MILLAR2022119228} FC-like measures derived from magnetoencephalography (MEG) have also been used for predicting age\cite{Xifra-Porxas2021-us} and sex\cite{Ott2021-cb}. Genomics such as single nucleotide polymorphisms (SNPs) can be used to make predictions that are much more accurate than those based on fMRI.\cite{Krishnan2016-cq}\cite{Liu2020-gx} Use of genomics and fMRI together may give superior results.\cite{Hu2021-df}

\begin{figure}
    \centering
    \includegraphics[width=12cm]{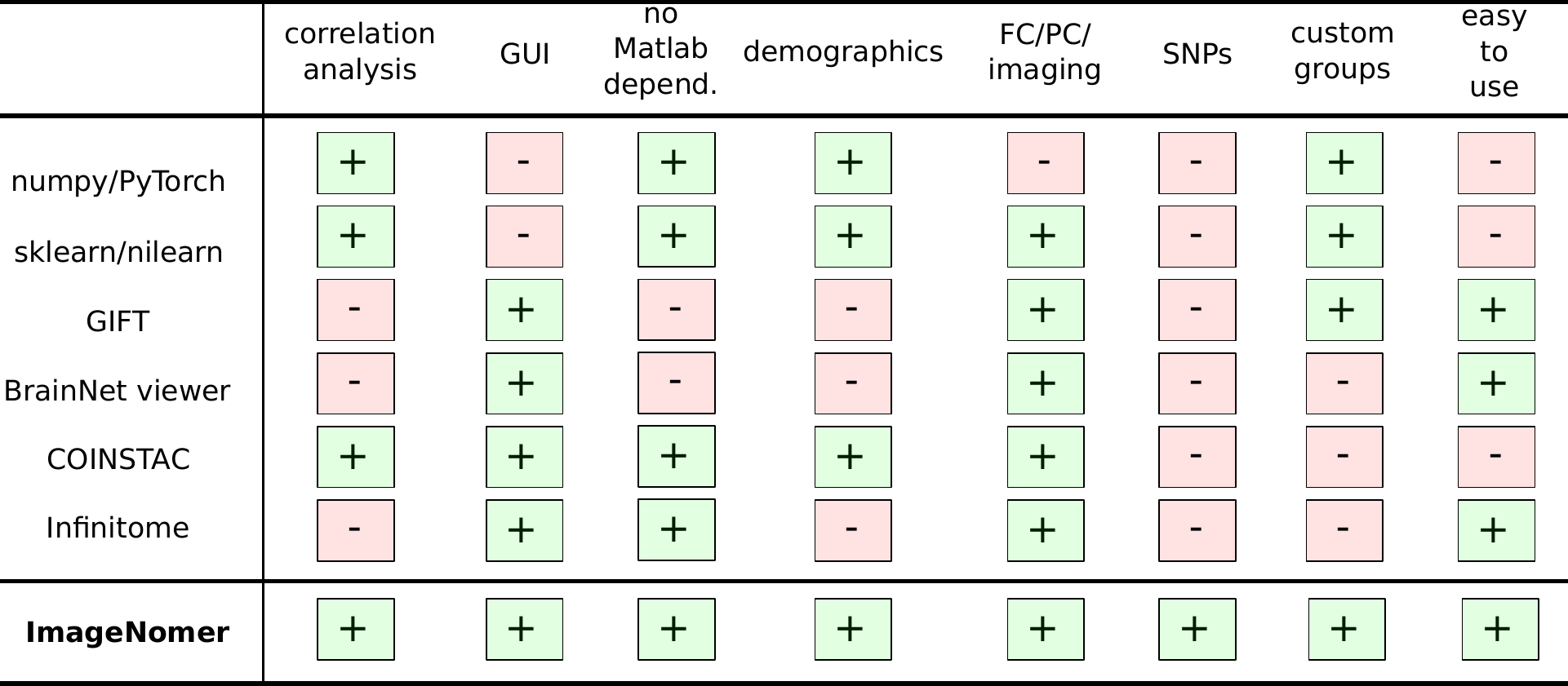}
    \caption{Comparison of existing toolkits for analysis of fMRI-based FC data with our ImageNomer software. A more comprehensive list may be found at \url{https://en.wikipedia.org/wiki/List_of_functional_connectivity_software}.}
    \label{fig:packages}
\end{figure}

Existing software packages for analysis of fMRI, FC, and FC-like measures such as partial correlation connectivity are either mostly text-based (programmatic interface) or have incomplete feature sets for identifying correlations in phenotypes (see Figure~\ref{fig:packages}). For example, numpy,\cite{harris2020array} PyTorch,\cite{NEURIPS2019_9015} scikit-learn,\cite{scikit-learn} nilearn,\cite{Abraham2014} and nipype\cite{gorgolewski_2016_50186} are all powerful and popular Python-based toolkits that can be used to conduct neuroimaging research. In fact, we use some of these packages as components in our ImageNomer software, but they all lack a graphical user interface that can speed up exploration of new datasets.  Classic packages such as the Matlab-based BrainNet viewer\cite{Xia2013} or GIFT toolbox,\cite{Calhoun2001} although they have a GUI frontend, do not allow for analysis of correlations between phenotypes as well as between phenotypes and imaging features/SNPs. Additionally, a Matlab-based toolchain ties one's product to a proprietary and non-free dependency. Even more modern tools like COINSTAC\cite{Plis2016} fall short because of a complicated user interface, lack of support for extremely high dimensional features, and a focus on federated learning which most neuroscientists do not need in their research. In contrast, ImageNomer focuses on data exploration by allowing correlation analysis of imaging, demographic, and genomic features and the creation of demographic-based subgroups. An overview of the ImageNomer architecture is shown in Figure~\ref{fig:architecture}.

Two problems with creating good, easy-to-use tools for analysis of fMRI-derived FC data are the high dimensionality of imaging features and the small effect sizes being measured. For example, Bennet et al. found that many effects found as marginally significant by standard analysis techniques are simply due to noise.\cite{Lyon2017-rd}\cite{Bennett2010-gy} For many recent fMRI studies, high dimensionality of the data and small effect size is exacerbated by small cohort sizes,\cite{SZUCS2020117164} with the average reproducible cohort size for an fMRI result being 36 subjects.\cite{Turner2018-vq} Our ImageNomer software addresses these points by treating visualized FC matrices (see Figure~\ref{fig:fcpanel}) and FC/phenotype correlation maps (see, e.g., Figure~\ref{fig:nullcorr}) as the primary outcome of the analysis, allowing quick visual inspection of what would take a long time through a programming interface. Cognizant of the high dimensionality of imaging features, we also perform Bonferroni-type multiple comparison correction in all FC-phenotype and SNP-phenotype correlation analysis. This does a lot to avoid the dead-salmon effect found by Bennet et al.\cite{Bennett2010-gy}



To demonstrate the utility of our developed ImageNomer tool, we use its data visualization and correlation abilities to quickly and easily identify a race confound in FC data. Specifically, we find that the high correlation between FC and race and the unequal distribution of achievement scores among races makes it appear that FC can predict achievement score, when our work shows it is more likely due to a confound. Many studies have used FC features to predict scholastic achievement, as measured by, e.g., WRAT score,\cite{Sayegh2014QualityOE} explaining 10\% of the variance in a population\cite{10002422} or achieving a small correlation with ground truth of $\rho\approx0.3$.\cite{Pervaiz2020-fm} We show, however, that the FC feature to WRAT score correlation is probably due to a confounding effect of race on FC. Indeed, previous studies have shown that AI models can sometimes trivially detect and be confounded by race,\cite{Gichoya2022-kt} and recent work has suggested that race can confound FC-based prediction of behavior.\cite{Li2022-rd} In this work, we use ImageNomer to identify a confound in FC, and find that this race confound is primarily responsible for any ability to predict WRAT score from FC. The utility of a tool like ImageNomer is validated by speed with which we find the race confound using a GUI toolkit, while numerous groups continue to search for achievement-based features, presumably using programmatic interfaces.\cite{10002422}\cite{Pervaiz2020-fm}



In summary, correlation analysis can give a quick overview of the data, and subject-level or cohort-level views can be instrumental for quality control. This is the reason we have developed ImageNomer, a visualization and analysis tool for connectivity-based fMRI and omics studies. The tool enables rapid correlation analysis as well as the comparison of features from outside models in a convenient browser-based user interface. Additionally, we include the ability to analyze distribution of phenotypes. Indeed, we find that correlation analysis is sufficient to quickly and clearly identify the confounding effect of race on WRAT score found in our study. Our code, as well as a Docker image and a live on-line demo, has been released and are available via links on our GitHub page.\footnote{Available online at \url{https://github.com/TulaneMBB/ImageNomer}}

\section{METHODS}

\subsection{Architecture}

\begin{figure} 
    \begin{center}
    \begin{tabular}{c} 
    \includegraphics[width=16cm]{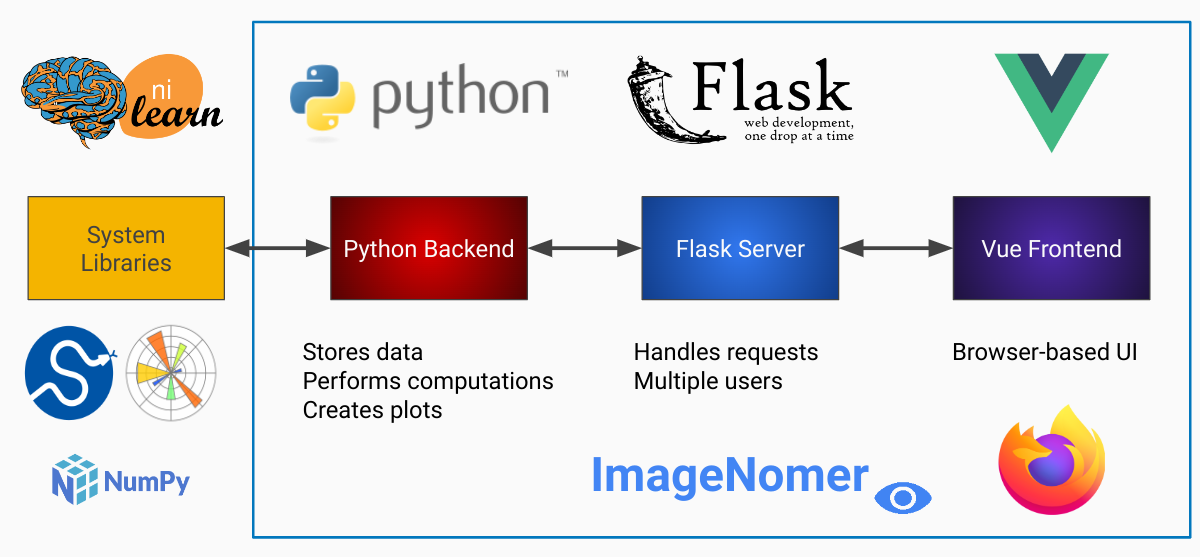}
    \end{tabular}
    \end{center}
    \caption[architecture] 
    { \label{fig:architecture} 
    Overview of the ImageNomer architecture.}
\end{figure} 

ImageNomer is made up of three components (see Figure~\ref{fig:architecture}):

\begin{itemize}
    \item a Python backend which integrates with available libraries such matplotlib, scikit-learn, and nilearn
    \item a Flask server that handles requests from the browser-based UI to the backend
    \item a Vue frontend which provides an interactive user experience from within the browser 
\end{itemize}

\begin{figure} 
    \begin{center}
    \begin{tabular}{c} 
    \includegraphics[width=17cm]{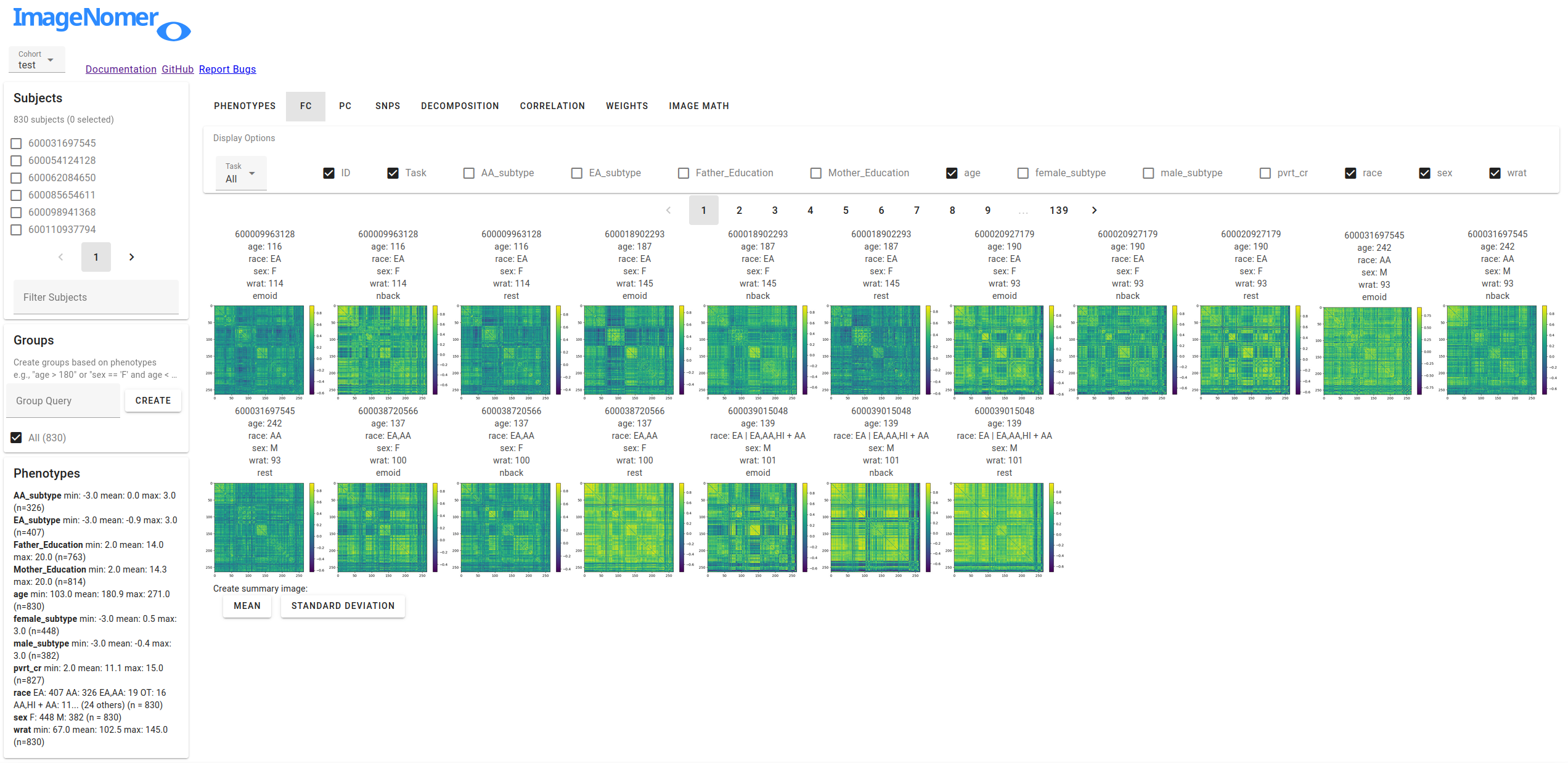}
    \end{tabular}
    \end{center}
    \caption[architecture] 
    { \label{fig:fcpanel} 
    Main view of the ImageNomer program showing resting state FC for all subjects along with demographic data.}
\end{figure}

A web-based user interface allows quick navigation around a cohort as well as the creation of summary graphs and correlation analyses. The main FC view is shown in Figure~\ref{fig:fcpanel}. The data being explored is stored locally in the server component, while the Python backend allows integration with standard libraries such as nilearn, scipy, numpy, and matplotlib. The matplotlib backend is used to generate all graphs on the backend, which are sent to the frontend as images. The Vue frontend allows for modularity of UI components, provides a library of pre-built widgets via Vuetify, and enables easy-to-code interactivity.

\subsection{Software features}

ImageNomer has the following capabilities:

\begin{itemize}
    \item Examine individual subject FC and partial correlation-based (PC) connectivity 
    \item Display distributions of phenotypes
    \item Correlate phenotypes with phenotypes, FC/PC features with phenotypes, and SNPs with phenotypes
    \item Display p-value maps for correlations
    \item Perform math on images
    \item Display components for FC decompositions (such as PCA)
    \item Correlate decomposition components with phenotypes or SNPs
    \item Display weights from machine learning models
    \item Summarize and average weights from multiple models
\end{itemize}

Future work with fMRI will likely require summarizing connectivity patterns into discrete network contributions.\cite{Wang2021-zs} In the future, we plan to expand ImageNomer's capabilities for summary measures and dictionary learning.

\subsection{Live web demo, Docker images, and tutorial}
\label{sec:links}

We have created a live on-line demo\footnote{On-line demo available at \url{https://aorliche.github.io/ImageNomer/live/}} and a Docker image containing an example open-access dataset of Fibromyalgia patients. This dataset is available as accession number ds004144 from OpenNeuro.\cite{ds004144:1.0.2}\cite{Markiewicz2021-wj} Instructions for using the Docker images, as well as a tutorial based around the Fibromyalgia dataset, can be found online.\footnote{Documentation and tutorial link: \url{https://imagenomer.readthedocs.io/en/latest/index.html}} The tutorial goes through step-by-step each of the major functions of ImageNomer, with instructions and screenshots of the expected output. Unfortunately, NIH data access policy precludes us from making the PNC or BSNIP data available publicly. If you are an approved researcher, we would be happy to work with you regarding functions, e.g., SNPs, which are not found in the Fibromyalgia dataset.

The easiest way to use ImageNomer is by mapping a directory on your local machine containing a ``demographics.pkl" file and an ``fc" subdirectory into the Docker image when starting the container.\footnote{Download and use ImageNomer: \url{https://github.com/TulaneMBB/ImageNomer}} We provide a second preprocessed OpenNeuro dataset ds004775\cite{ds004775:1.1.1} dealing with Vicarious Punishment in our GitHub along with a tutorial on how to map it to the Docker container, similar to what one would do for one's own dataset. Our GitHub repository also contains a notebooks folder with Jupyter notebooks that shows step-by-step how the Fibromyalgia and Punishment data were preprocessed in manner suitable for ImageNomer, starting from a CSV/TSV file and BOLD timeseries. 

Alternatively, ImageNomer can be used by cloning our GitHub project and installing the Python requirements via pip. However, the use of Docker images, via instructions found on our GitHub (\url{https://github.com/TulaneMBB/ImageNomer}) is the easiest method. Docker images have been built for the amd64 and arm64 architectures; check the documentation for how to use the right version for you. If one is interested in editing the code, it is split into the ``backend" and ``frontend" directories. The ``backend" directory contains Python modules and does the heavy lifting with respect to data loading, image generation, and correlation analysis. The ``frontend" directory contains a Vue javascript project that handles the browser-based interface and keeps track of most session state. Individual parts of the web-interface are built as Vue components. 

\subsection{Case study on ethnicity confound in FC and its impact on achievement score prediction}

As a demonstration of the power of ImageNomer's GUI in quickly identifying trends, confounds, and correlations in data, we give a case study of using ImageNomer to identify an under-reported race confound present in FC. As discussed in the Introduction, many groups have tried to predict achievement score or similar metrics from FC. Using ImageNomer, we find that there is a large difference between achievement scores among races. This can potentially lead to a confounding effect if race can be predicted from FC. We also find a high correlation between race and certain FC regions, making us suspect that race-from-FC prediction is possible. We then verify the presence of this confound by performing regression on the whole cohort vs. within-ethnicity subsets. We learn that quick and dirty data exploration may save a lot of time trying to look for FC correlates of cognition that may or may not be there.

\subsection{PNC dataset}

\begin{figure} 
    \begin{center}
    \begin{tabular}{c} 
    \includegraphics[width=17cm]{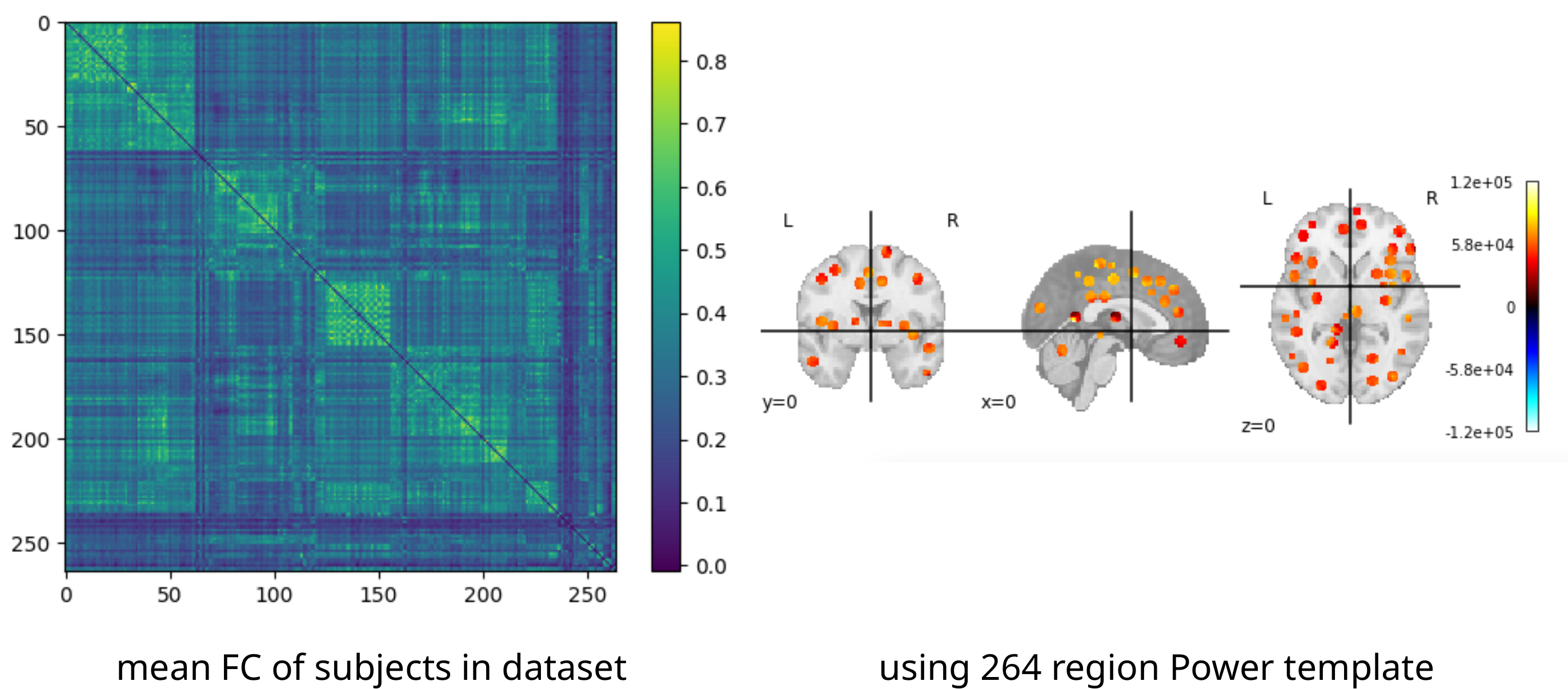}
    \end{tabular}
    \end{center}
    \caption[architecture] 
    { \label{fig:fcpower} 
    Mean FC in the PNC dataset along with the Power264 template\cite{Power2011FunctionalNO} regions used to sample BOLD signal from brain regions.}
\end{figure}

We tested ImageNomer by using it to examine the large Philadelphia Neurodevelopmental Cohort (PNC) dataset.\cite{Glessner2010-xg}\cite{Satterthwaite2014NeuroimagingOT} The PNC dataset contains fMRI scans, SNP information, cognitive batteries, questionnaires, and phenotype data from healthy adolescents between 8-23 years old. The dataset is enriched for European Ancestry (EA) and African Ancestry (AA) races. It contains fMRI scans for 1,445 healthy adolescents and SPN data for more than 9,267. We chose an 830-subject subset of the data which included subjects with SNP information as well as resting state (rest), working memory (nback), and emotion identification (emoid) scanner task fMRI scans. Scholastic achievement and problem-solving ability was measured by Wide Range Achievement Test (WRAT) score\cite{Sayegh2014QualityOE} with the effects of age regressed out. A total of three fMRI tasks were acquired: resting state, working memory, and emotion identification. An example of parcellation along with mean FC in the PNC dataset is shown in Figure~\ref{fig:fcpower}. The acquisition parameters for fMRI and FC preprocessing have been described elsewhere.\cite{10002422}





SNPs were collected using one of eight different platforms, with the largest set containing 1,185,051 SNPs \cite{Glessner2010-xg}. We selected a subset of 10,433 SNPs that were found in at least 100 subjects in the cohort for our analysis. SNPs were categorized by haplotype as homozygous minor variant, heterozygous, and homozygous major variant. Missing values for subjects were set to zero for all haplotypes.

\subsection{BSNIP dataset}

Robustness of race prediction from FC was tested by using an independent dataset to validate models trained on PNC. The dataset used was the Bipolar and Schizophrenia Network for Intermediate Phenotypes (BSNIP) dataset of 933 patients, 1059 relatives, and 459 healthy controls.\cite{Tamminga2014-pp} fMRI scans were acquired over 6 different sites, and acquisition and preprocessing are described elsewhere.\cite{8857902} For validation of race prediction we chose a subset of 387 African Americans (AA) and 778 Caucasians (CA), both patients and healthy controls, for whom we had fMRI scans.






\section{RESULTS AND DISCUSSION}

We first present our exploration of the potential race confound in achievement score prediction from FC using ImageNomer. Based on analysis with ImageNomer, we hypothesize that due to the high correlation of FC with race and the obvious difference in achievement scores between races, prediction of achievement score from FC is solely due to a race confound. We then corroborate our hypothesis by using whole cohort and within-ethnic group regression models. Note that all Figures presented in this section are screenshots from the ImageNomer program.

\subsection{Data exploration of possible race confound with ImageNomer}

\begin{figure}
    \begin{center}
    \begin{tabular}{c} 
    \includegraphics[width=17cm]{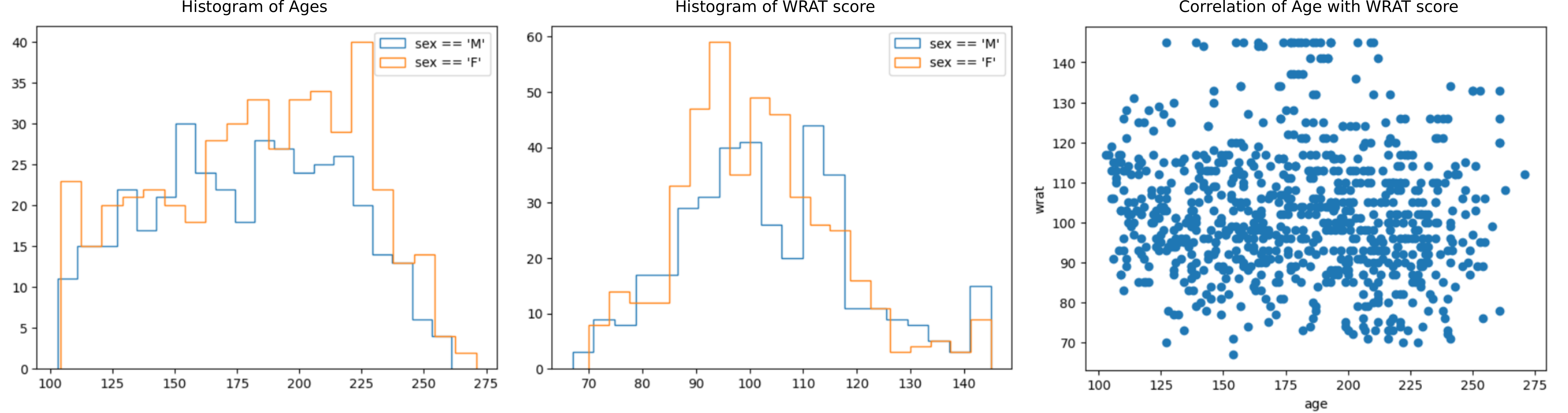}
    \end{tabular}
    \end{center}
    \caption[architecture] 
    { \label{fig:demo} 
    Demographics of our subset of the PNC dataset. Plots of age vs sex, WRAT vs sex, and WRAT vs age are shown. All plots created using the GUI of ImageNomer, without programming input.}
\end{figure}

\begin{figure} 
    \begin{center}
    \begin{tabular}{c} 
    \includegraphics[width=17cm]{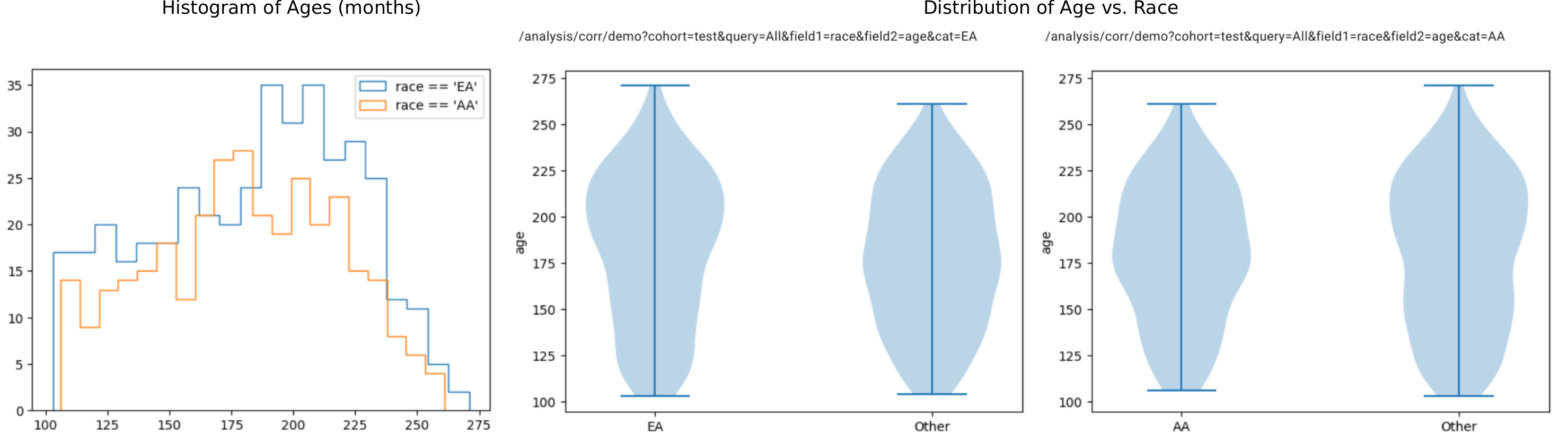}
    \end{tabular}
    \end{center}
    \caption[architecture] 
    { \label{fig:agebias} 
    Distribution of race vs age. We see that there is no race bias on age distribution.}
\end{figure}

We first confirm that age and sex are not possible confounding factors with respect to achievement score prediction. This is illustrated in Figures~\ref{fig:demo} and \ref{fig:agebias}, where we see equal distributions of WRAT score among males and females and no correlation with age (raw WRAT score has been corrected for age).

\begin{figure}
    \begin{center}
    \begin{tabular}{c} 
    \includegraphics[width=12cm]{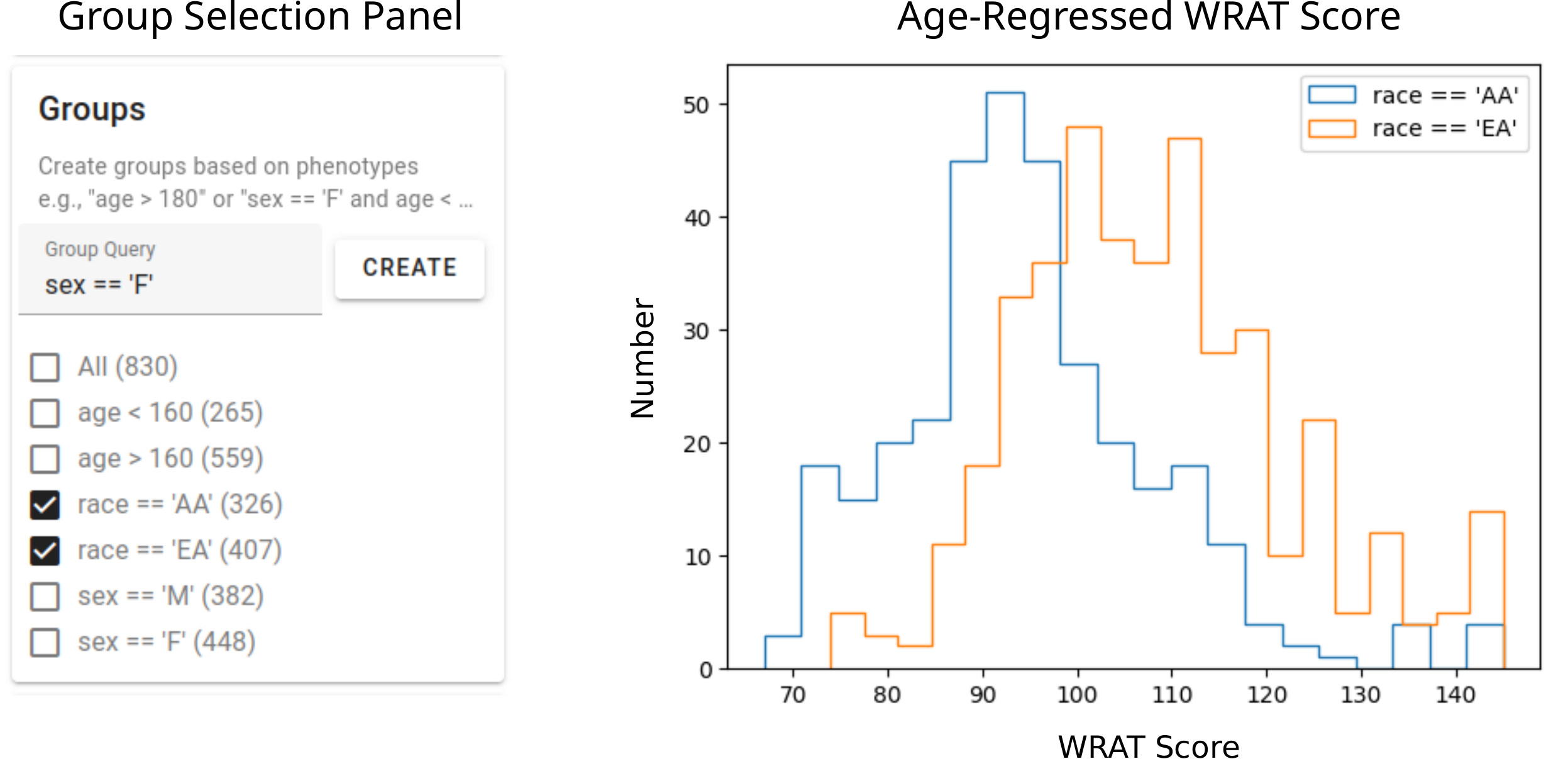}
    \end{tabular}
    \end{center}
    \caption[architecture] 
    { \label{fig:bias} 
    Examining race bias on WRAT score in the PNC dataset using ImageNomer. We find whereas age has been regressed from WRAT score, there is still a large racial bias.}
\end{figure}

Next, we use the group creation capabilities of ImageNomer to create two groups: European Ancestry (EA) and African Ancestry (AA) groups. We then compare the WRAT score distribution between the two groups, illustrated in Figure~\ref{fig:bias}. We find that here there is a clear difference in achievement score distribution, leading to the possibility of a confounding effect if there is a race signal present in FC.

\begin{figure} 
    \begin{center}
    \begin{tabular}{c} 
    \includegraphics[width=17cm]{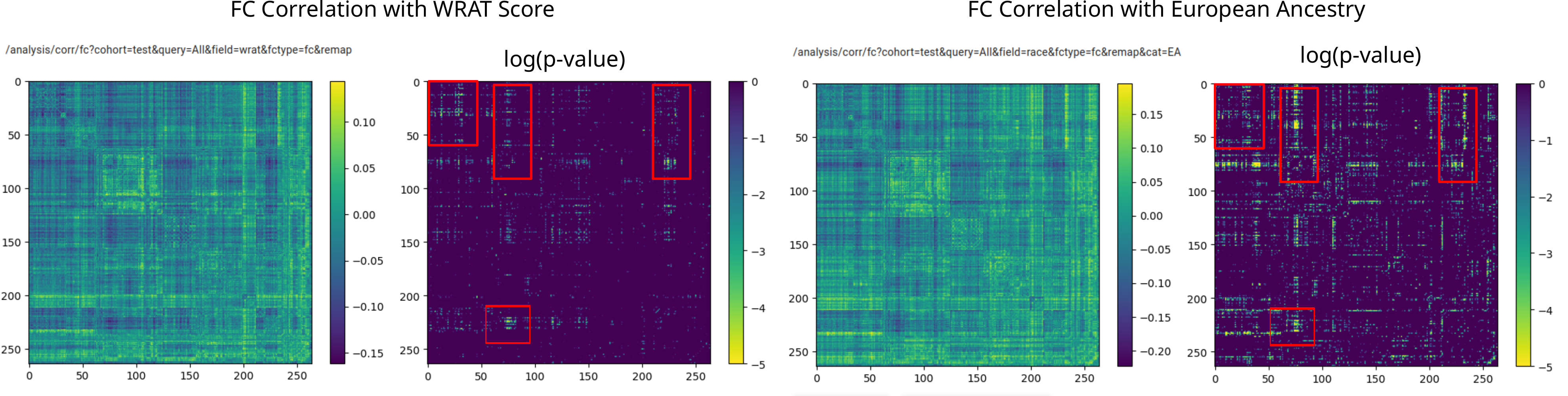}
    \end{tabular}
    \end{center}
    \caption[architecture] 
    { \label{fig:nullcorr} 
     Correlation between race and FC is much higher than the correlation between race and WRAT score. Additionally, in almost all regions, achievement score-correlated FC is a subset of race-correlated FC.}
\end{figure}

Using the FC-to-phenotype correlation feature of ImageNomer, we explore whether there is correlation between race and FC. In Figure~\ref{fig:nullcorr}, we find that there is a large and significant correlation between race and FC. Furthermore, in the same Figure, we show that the smaller correlation between WRAT score and FC is actually a subset of the race-FC correlation.

\begin{figure}
    \centering
    \includegraphics[width=17cm]{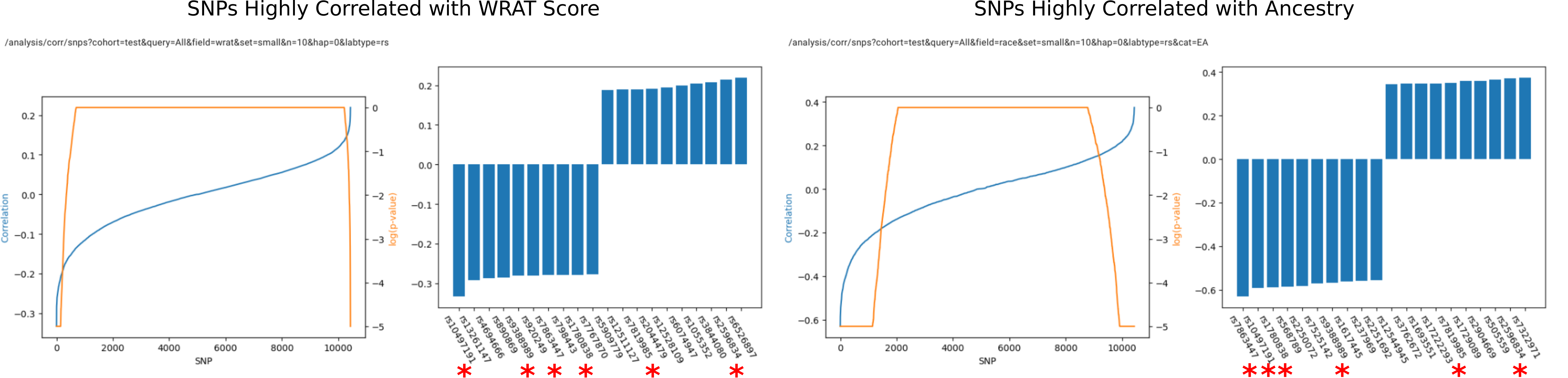}
    \caption{ImageNomer identifies a large overlap between SNPs correlated with race and SNPs correlated with WRAT score. Out of the top 20, 6 of the same SNP appear in both groups (marked with red asterisks).}
    \label{fig:snpscorr}
\end{figure}

We perform the same analysis for SNPs, with the caveat that sex and race can be perfectly predicted using SNPs, and that SNP data does not contain any age-related signal. Nevertheless, we still attempted to find whether there was a suggestive overlap between SNPs correlated with race and SNPs correlated with high or low WRAT score. Our results are shown in Figure~\ref{fig:snpscorr}. From a total of 10,433 SNPs found in 100 or more subjects, we identified the top 20 SNPs correlated with race and achievement score. Of these top 20, six appeared in both the highly WRAT-correlated and highly race-correlated batches.

In summary, we use ImageNomer to form the hypothesis that race may bias FC-based prediction of achievement score if there is a race signal present in FC. We then find, using ImageNomer, that the race signal present in FC is in fact stronger than the signal for achievement score, and that WRAT score to FC correlation is a subset of race-FC correlation. Finally, we draw the same conclusion in SNP to race and SNP to achievement score correlation. In the next section, we describe the use of regression models to validate our hypothesis.

\subsection{Validation with regression models}

We validate the qualitative results from ImageNomer's data visualization and exploration capabilities with train/test regression models. We use regularized Ridge and Logistic Regression models with an 80/20 train-test split and 20 bootstrapping repetitions to predict age, sex, race, and WRAT score. Additionally, we predict WRAT score in whole cohort as well as within intra-ethnicity groups. The results are shown in Table~\ref{tab:predsummary}.

The results are as follows: age, sex, and race can all be modestly well predicted using FC. WRAT score can be predicted, although at a barely significant level, using the whole cohort, with both FC and/or SNPs as input. However, any ability to predict WRAT score disappears in race-controlled (within ethnicity) groups. This validates our hypothesis, formulated with ImageNomer via data exploration, that FC features used to predict achievement score are actually predicting ethnicity instead.

Next, we confirm the stability of race signal in FC by using both ImageNomer and transfer learning of regression models to find that race signal is at least somewhat conserved between the PNC and BSNIP datasets. Finally, we consider the effect of socioeconomic status (SES) as another potential confound besides race in predicting achievement score from FC. 



\begin{table}
\begin{center}
\begin{tabular}{|l|c|c|c|c|c|} 
\hline
Prediction & Modality & Metric & Null Model & Best Full Model & Best 10 Features\\
\hline
\hline
Age & FC & RMSE, months & 38.4 & \textbf{26} & \textbf{32.2} \\
\hline
WRAT Score & FC & RMSE & 15.1 & \textbf{13.6} & 15.1 \\
\hline
WRAT Score & SNPs & RMSE & 15.1 & \textbf{14} & - \\
\hline
WRAT Score (AA) & FC & RMSE & 13.9 & 13.8 & 13.9 \\
\hline
WRAT Score (AA) & SNPs & RMSE & 13.9 & 13.4 & - \\
\hline
WRAT Score (EA) & FC & RMSE & 14 & 14.1 & 14 \\
\hline
WRAT Score (EA) & SNPs & RMSE & 14 & 13.6 & - \\
\hline
Race & FC & Accuracy & 55\% & \textbf{85\%} & \textbf{72\%} \\
\hline
Sex & FC & Accuracy & 51\% & \textbf{78\%} & \textbf{62\%} \\
\hline
\end{tabular}
\end{center}
\caption{
\label{tab:predsummary}Summary of prediction results for full models (34,716 features for FC, 10,433 features for SNPs) and top 10 feature models in the PNC dataset. Top 10 features selected on the training set. Statistically significant results are shown in bold.}
\end{table}

\subsubsection{Transfer of race prediction models between PNC and BSNIP datasets}

\begin{figure}
    \begin{center}
    \begin{tabular}{c} 
    \includegraphics[width=16cm]{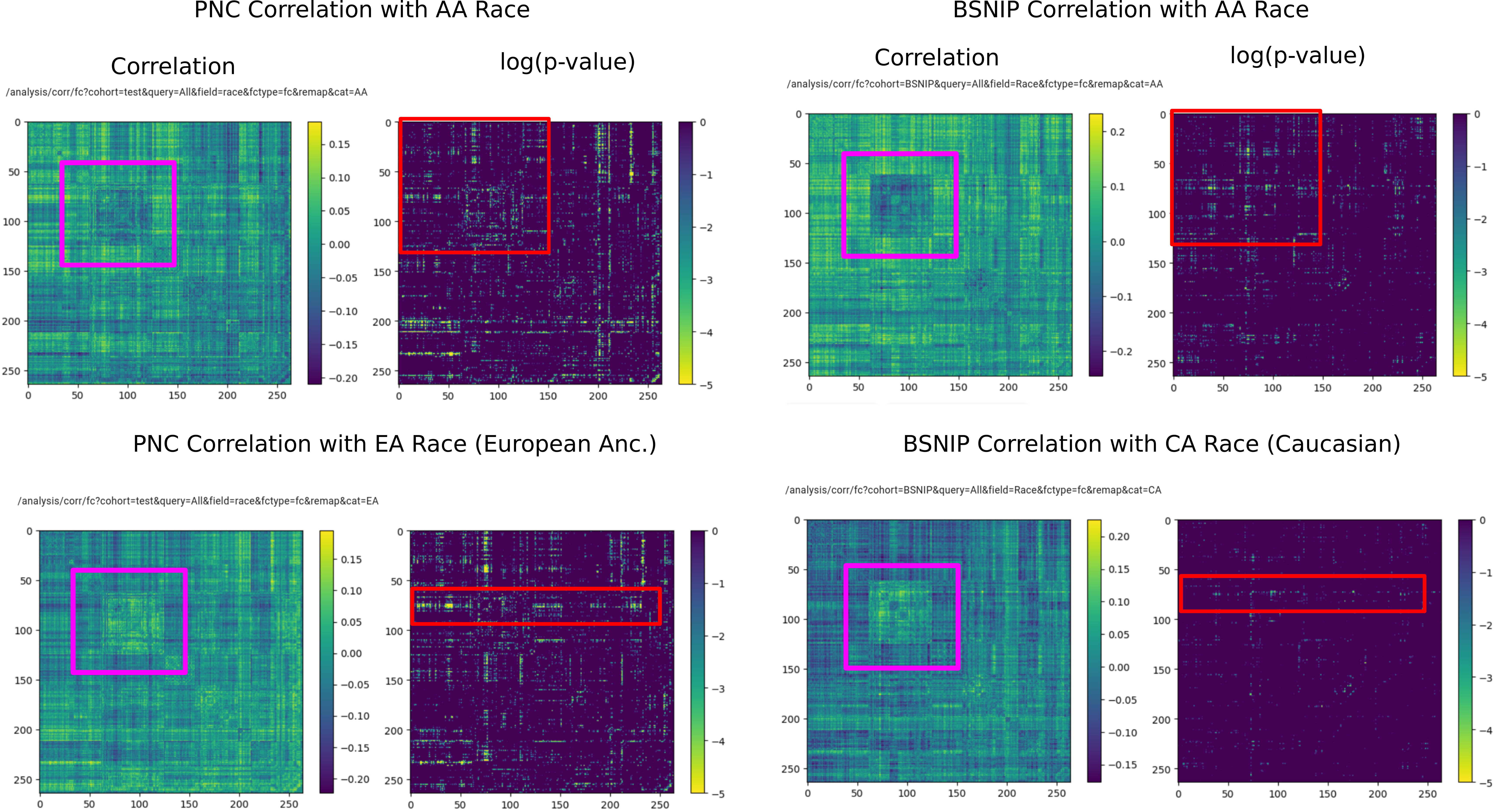}
    \end{tabular}
    \end{center}
    \caption[architecture] 
    { \label{fig:transfer} 
    Correlation of FC with race in the PNC and BSNIP datasets. Although BSNIP p-values are lower, we see the same pattern of correlation. Additionally, the overall FC in the Default Mode Network (DMN), highlighted in pink, seems to be highly predictive of race.}
\end{figure}

We show screenshots of ImageNomer-based data exploration for FC correlation with race in the PNC and BSNIP datasets in Figure~\ref{fig:transfer}. This figure highlights the fact that both datasets have similar correlations between specific FCs and race. We confirm the ImageNomer-based hypothesis with results for transfer of race prediction models between the PNC and BSNIP datasets, shown in Table~\ref{tab:transfer}. A Logistic Regression model trained on the PNC dataset was able to predict race in the BSNIP dataset with an average accuracy of 68\%. When trained on BSNIP and evaluated on PNC, the average prediction accuracy was 66\%. We find that the prediction is less good than within-dataset prediction, although still better than chance. It should be taken into account that the PNC dataset is made up of healthy adolescents, while the BSNIP dataset contains schizophrenia and bipolar patients, relatives of patients, and healthy controls. Figure~\ref{fig:transfer} shows a comparison between race correlation and FC in the PNC and BSNIP datasets, created using ImageNomer.

\begin{table}
\begin{center}
\begin{tabular}{|c|c|} 
\multicolumn{2}{c}{} \\
\hline
\multicolumn{2}{|c|}{Trained on PNC} \\
\hline
\hline
Evaluation Group & Accuracy \\
\hline 
PNC (all, n=733) & 85$\pm$3\% \\ 
\hline 
BSNIP AA (n=387) & 76$\pm$5\% \\
\hline 
BSNIP CA (n=778) & 64$\pm$5\% \\
\hline
\end{tabular}
\begin{tabular}{|c|c|} 
\multicolumn{2}{c}{} \\
\hline
\multicolumn{2}{|c|}{Trained on BSNIP} \\
\hline
\hline
Evaluation Group & Accuracy \\
\hline
BSNIP (all, n=1165) & 79$\pm$4\% \\
\hline 
PNC EA (n=407) & 90$\pm$3\% \\
\hline 
PNC AA (n=326) & 38$\pm$7\% \\
\hline
\end{tabular}
\end{center}
\caption{
\label{tab:transfer}Accuracy of transfer learning between the PNC and BSNIP datasets. All predictions are better than the null model, except for identification of the AA group in the PNC dataset by a model trained on BSNIP.}
\end{table}

\subsection{Effect of socioeconomic status (SES) explored with ImageNomer and regression models}

We consider socioeconomic status (SES) as another confounding factor when predicting scholastic achievement based on a standardized test, with the majority of analysis again carried out using ImageNomer. Predictive models were only used to validate the conclusions made using data exploration in ImageNomer.

\begin{figure}
    \begin{center}
    \begin{tabular}{c} 
    \includegraphics[width=17cm]{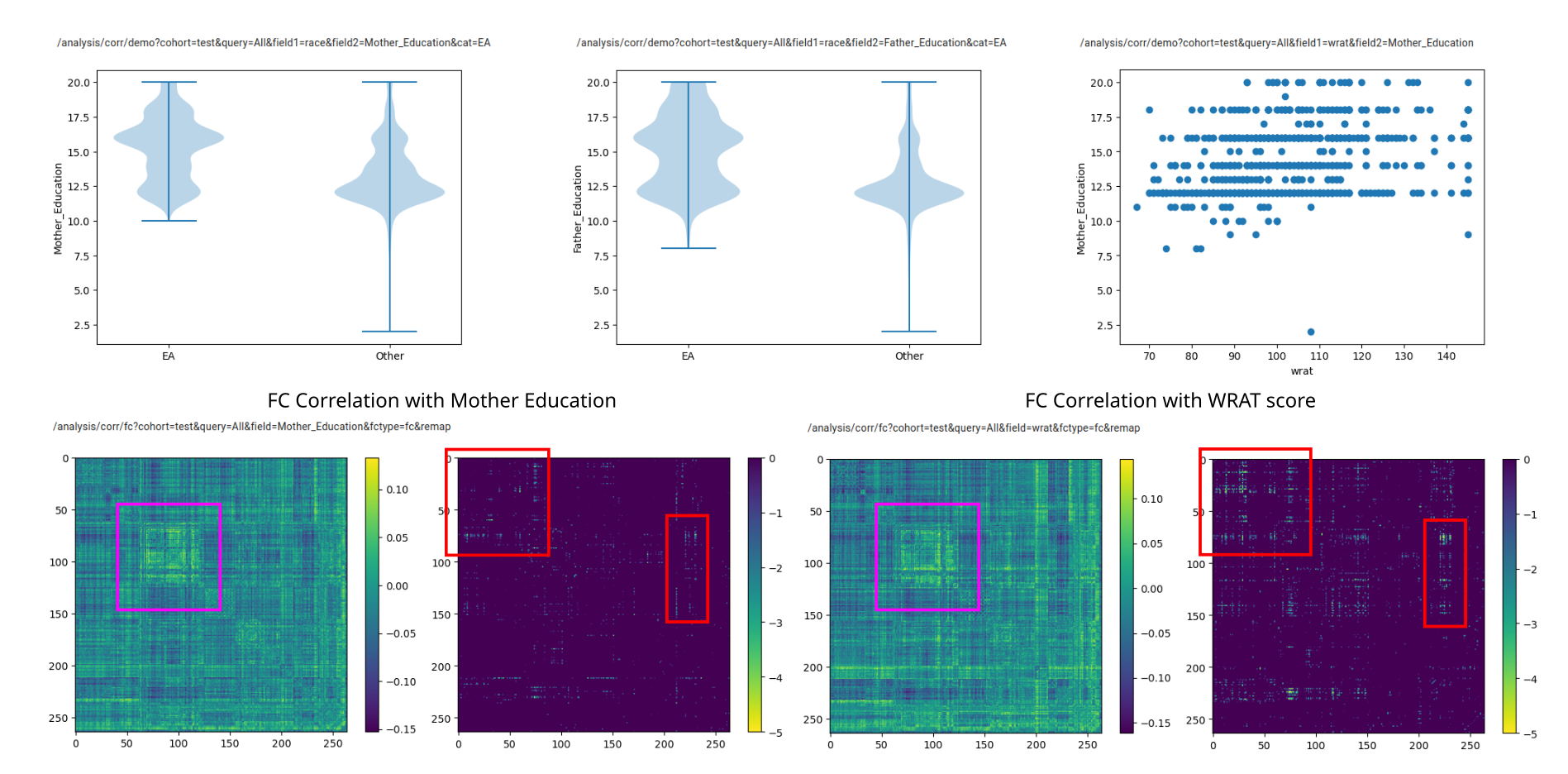}
    \end{tabular}
    \end{center}
    \caption
    { \label{fig:ses} 
    Top: distribution of mother and father education level by race category and distribution of WRAT score with mother education level. Bottom: correlation of FC with mother education level and WRAT score. Correlation consists of a correlation image and the negative log base 10 of the Bonferroni-corrected p-value, clipped at -5.}
\end{figure}

A problem is that SES was not directly measured in the PNC study, in that the income of family groups was not known. However, previous studies have used parental education levels as a proxy for SES,\cite{Chen2018-py} and this information was included in the PNC dataset. Indeed, as seen in Figure~\ref{fig:ses}, we find that SES, race, and WRAT score are all inter-related. We see that non-EA ethnicity tend to have lower SES as measured by mother education level. The correlation of father education level with FC was similar to mother education, although less significant. It should be noted that many children had missing values for father education level.

\begin{table}
\begin{center}
\begin{tabular}{|c|c|c|} 
\multicolumn{3}{c}{}\\
\hline
Group & WRAT score RMSE & Null model RMSE \\
\hline
Low SES (Mother Education $\leq 12$ years) & 14.1 & 15 \\ 
\hline
High SES (Mother Education $\geq 14$ years) & 15 & 15.3 \\
\hline
\end{tabular}
\end{center}
\caption{
\label{tab:ses}Prediction of WRAT score in the low SES (mother with no college education) versus high SES (mother with some college education). Predictive ability was barely significant in the low SES group, and not significant in the high SES group.}
\end{table}

We see in Figure~\ref{fig:ses} that SES, ethnicity, and WRAT score correlate to similar regions on the FC map. The p-values associated with FC-SES correlation (as measured by mother education) are somewhat lower than those associated with FC-WRAT or FC-ethnicity correlation, but still significant. Additionally, performing regression analysis for WRAT score based on FC in low SES (mother education $\leq 12$ years) and high SES (mother education $\geq 14$ years) groups, we find a barely significant ability to predict achievement in the low SES group but not a significant ability in the high SES group. The results are shown in Table~\ref{tab:ses}. The WRAT prediction accuracy in the low SES group is worse than in the cohort as a whole (compare 13.6 RMSE vs 14.1 RMSE). We conclude that SES is a confounding factor in WRAT score prediction, though not as severe as ethnicity in this dataset. Finally, we reject the idea that race is a causal factor in achievement or WRAT score, but only point out the potential confounding effects if race or SES is not taken into account in studies seeking to find markers of high or low achievement, and the ease with which such confounds were found using ImageNomer.

\section{CONCLUSION}

We present ImageNomer, a new fMRI and omics visualization and analysis tool. We note that most of the figures shown in this manuscript were created as screenshots of the working tool. We use this tool to examine the large PNC dataset and discover features important for phenotype prediction. As validation for ImageNomer-based correlation analysis, we find that age, sex, and race can be moderately well predicted by FC features, with 10 FC features giving up to 72\% race prediction accuracy, compared with 85\% for the full model.

We find both FC features and SNPs can somewhat predict scholastic knowledge and problem-solving ability, as measured by WRAT score, but that this is probably due to a race confound. When controlling for race, FC-based achievement score prediction drops to the same accuracy as the null model and the SNP-based prediction becomes statistically insignificant. We conclude that, on average, the effect of either SNPs or FC features on scholastic achievement in normal children is very small, if one exists at all. Additionally, we find that race prediction from FC is at least somewhat robust between different datasets. Using ImageNomer, this work quickly and easily identifies race as an important confounding factor in FC and casts doubt on the ability to predict achievement-related features from both FC and SNP data.

Finally, we note that it is very easy to add additional datasets to explore into the ImageNomer program. To do so, follow the links given in the footnotes in Section~\ref{sec:links} and read the corresponding instructions. Doing so requires following a Jupyter notebook, but once data is loaded into ImageNomer, it can be explored without writing any additional code. We find the ability to quickly visualize trends, correlations, and potential confounding effects provided by the ImageNomer software is invaluable to the ability to perform good and careful research. This is demonstrated by the rapid identification of a race confound on FC-based prediction of achievement using ImageNomer, despite the fact that many studies have attempted to predict achievement using FC with little or no mention of this effect.\cite{Sayegh2014QualityOE}\cite{10002422}\cite{Pervaiz2020-fm}

\acknowledgments 

The authors would like acknowledge the NIH (grants R01 GM109068, R01 MH104680, R01 MH107354, P20 GM103472, R01 EB020407, R01 EB006841, R56 MH124925, 5U19 AG055373), NSF (\#1539067), and American Heart Association (\#830166) for partial funding support.

PNC fMRI, SNP, and phenotype data came from the Neurodevelopmental Genomics: Trajectories of Complex Phenotypes database of genotypes and phenotypes repository, dbGaP Study Accession ID phs000607.v3.p2.

BSNIP data came from the National Institutes of Mental Health (NIMH) Data Archive (NDA).

Part of the work on ImageNomer was conducted at the UBRITE Multiomics Hackathon\footnote{\url{https://hackathon.ubrite.org/}} sponsored by the University of Alabama at Birmingham. The authors would like to acknowledge the organizers and mentors at UAB for their help in fostering teamwork and innovation in the bioinformatics community.

\bibliography{ImageNomer} 

\begin{thebibliography}{10}

\bibitem{Belliveau1991FunctionalMO}
Belliveau, J.~W., Kennedy, D.~N., McKinstry, R.~C., Buchbinder, B.~R., Weisskoff, R.~M., Cohen, M.~S., Vevea, J.~M., Brady, T.~J., and Rosen, B.~R., ``Functional mapping of the human visual cortex by magnetic resonance imaging.,'' {\em Science}~{\bf 254 5032},  716--9 (1991).

\bibitem{doi:10.1073/pnas.0135058100}
Greicius, M.~D., Krasnow, B., Reiss, A.~L., and Menon, V., ``Functional connectivity in the resting brain: A network analysis of the default mode hypothesis,'' {\em Proceedings of the National Academy of Sciences}~{\bf 100}(1),  253--258 (2003).

\bibitem{10002422}
Orlichenko, A., Qu, G., Zhang, G., Patel, B., Wilson, T.~W., Stephen, J.~M., Calhoun, V.~D., and Wang, Y.-P., ``Latent similarity identifies important functional connections for phenotype prediction,'' {\em IEEE Transactions on Biomedical Engineering} ,  1--12 (2022).

\bibitem{Zhang2020GenderDA}
Zhang, X., Liang, M., Qin, W., Wan, B., Yu, C., and Ming, D., ``Gender differences are encoded differently in the structure and function of the human brain revealed by multimodal mri,'' {\em Frontiers in Human Neuroscience}~{\bf 14} (2020).

\bibitem{Qu2021EnsembleMR}
Qu, G., Xiao, L., Hu, W., Wang, J., Zhang, K., Calhoun, V.~D., and ping Wang, Y., ``Ensemble manifold regularized multi-modal graph convolutional network for cognitive ability prediction,'' {\em IEEE Transactions on Biomedical Engineering}~{\bf 68},  3564--3573 (2021).

\bibitem{Du2018ClassificationAP}
Du, Y., Fu, Z., and Calhoun, V.~D., ``Classification and prediction of brain disorders using functional connectivity: Promising but challenging,'' {\em Frontiers in Neuroscience}~{\bf 12} (2018).

\bibitem{MILLAR2022119228}
Millar, P.~R., Luckett, P.~H., Gordon, B.~A., and Benzinger, T.~L., ``Predicting brain age from functional connectivity in symptomatic and preclinical alzheimer disease,'' {\em NeuroImage}~{\bf 256},  119228 (2022).

\bibitem{Xifra-Porxas2021-us}
Xifra-Porxas, A., Ghosh, A., Mitsis, G.~D., and Boudrias, M.-H., ``Estimating brain age from structural {MRI} and {MEG} data: Insights from dimensionality reduction techniques,'' {\em Neuroimage}~{\bf 231},  117822 (May 2021).

\bibitem{Ott2021-cb}
Ott, L.~R., Penhale, S.~H., Taylor, B.~K., Lew, B.~J., Wang, Y.-P., Calhoun, V.~D., Stephen, J.~M., and Wilson, T.~W., ``Spontaneous cortical {MEG} activity undergoes unique age- and sex-related changes during the transition to adolescence,'' {\em Neuroimage}~{\bf 244},  118552 (Dec. 2021).

\bibitem{Krishnan2016-cq}
Krishnan, A., Zhang, R., Yao, V., Theesfeld, C.~L., Wong, A.~K., Tadych, A., Volfovsky, N., Packer, A., Lash, A., and Troyanskaya, O.~G., ``Genome-wide prediction and functional characterization of the genetic basis of autism spectrum disorder,'' {\em Nat. Neurosci.}~{\bf 19},  1454--1462 (Nov. 2016).

\bibitem{Liu2020-gx}
Liu, Y., Xu, L., Li, J., Yu, J., and Yu, X., ``Attentional connectivity-based prediction of autism using heterogeneous {rs-fMRI} data from {CC200} atlas,'' {\em Exp. Neurobiol.}~{\bf 29},  27--37 (Feb. 2020).

\bibitem{Hu2021-df}
Hu, W., Meng, X., Bai, Y., Zhang, A., Qu, G., Cai, B., Zhang, G., Wilson, T.~W., Stephen, J.~M., Calhoun, V.~D., and Wang, Y.-P., ``Interpretable multimodal fusion networks reveal mechanisms of brain cognition,'' {\em IEEE Trans. Med. Imaging}~{\bf 40},  1474--1483 (May 2021).

\bibitem{harris2020array}
Harris, C.~R., Millman, K.~J., van~der Walt, S.~J., Gommers, R., Virtanen, P., Cournapeau, D., Wieser, E., Taylor, J., Berg, S., Smith, N.~J., Kern, R., Picus, M., Hoyer, S., van Kerkwijk, M.~H., Brett, M., Haldane, A., del R{\'{i}}o, J.~F., Wiebe, M., Peterson, P., G{\'{e}}rard-Marchant, P., Sheppard, K., Reddy, T., Weckesser, W., Abbasi, H., Gohlke, C., and Oliphant, T.~E., ``Array programming with {NumPy},'' {\em Nature}~{\bf 585},  357--362 (Sept. 2020).

\bibitem{NEURIPS2019_9015}
Paszke, A., Gross, S., Massa, F., Lerer, A., Bradbury, J., Chanan, G., Killeen, T., Lin, Z., Gimelshein, N., Antiga, L., Desmaison, A., Kopf, A., Yang, E., DeVito, Z., Raison, M., Tejani, A., Chilamkurthy, S., Steiner, B., Fang, L., Bai, J., and Chintala, S., ``Pytorch: An imperative style, high-performance deep learning library,'' in [{\em Advances in Neural Information Processing Systems 32}{\nolinebreak\hspace{0.1em}]},   8024--8035, Curran Associates, Inc. (2019).

\bibitem{scikit-learn}
Pedregosa, F., Varoquaux, G., Gramfort, A., Michel, V., Thirion, B., Grisel, O., Blondel, M., Prettenhofer, P., Weiss, R., Dubourg, V., Vanderplas, J., Passos, A., Cournapeau, D., Brucher, M., Perrot, M., and Duchesnay, E., ``Scikit-learn: Machine learning in {P}ython,'' {\em Journal of Machine Learning Research}~{\bf 12},  2825--2830 (2011).

\bibitem{Abraham2014}
Abraham, A., Pedregosa, F., Eickenberg, M., Gervais, P., Mueller, A., Kossaifi, J., Gramfort, A., Thirion, B., and Varoquaux, G., ``Machine learning for neuroimaging with scikit-learn,'' {\em Frontiers in Neuroinformatics}~{\bf 8} (2014).

\bibitem{gorgolewski_2016_50186}
Gorgolewski, K.~J., Esteban, O., Burns, C., Ziegler, E., Pinsard, B., Madison, C., Waskom, M., Ellis, D.~G., Clark, D., Dayan, M., Manhães-Savio, A., Notter, M.~P., Johnson, H., Dewey, B.~E., Halchenko, Y.~O., Hamalainen, C., Keshavan, A., Clark, D., Huntenburg, J.~M., Hanke, M., Nichols, B.~N., Wassermann, D., Eshaghi, A., Markiewicz, C., Varoquaux, G., Acland, B., Forbes, J., Rokem, A., Kong, X.-Z., Gramfort, A., Kleesiek, J., Schaefer, A., Sikka, S., Perez-Guevara, M.~F., Glatard, T., Iqbal, S., Liu, S., Welch, D., Sharp, P., Warner, J., Kastman, E., Lampe, L., Perkins, L.~N., Craddock, R.~C., Küttner, R., Bielievtsov, D., Geisler, D., Gerhard, S., Liem, F., Linkersdörfer, J., Margulies, D.~S., Andberg, S.~K., Stadler, J., Steele, C.~J., Broderick, W., Cooper, G., Floren, A., Huang, L., Gonzalez, I., McNamee, D., Papadopoulos~Orfanos, D., Pellman, J., Triplett, W., and Ghosh, S., ``{Nipype: a flexible, lightweight and extensible neuroimaging data processing framework in Python. 0.12.0-rc1},'' (Apr.
  2016).

\bibitem{Xia2013}
Xia, M., Wang, J., and He, Y., ``{BrainNet} viewer: A network visualization tool for human brain connectomics,'' {\em {PLoS} {ONE}}~{\bf 8},  e68910 (July 2013).

\bibitem{Calhoun2001}
Calhoun, V., Adali, T., Pearlson, G., and Pekar, J., ``A method for making group inferences from functional {MRI} data using independent component analysis,'' {\em Human Brain Mapping}~{\bf 14}(3),  140--151 (2001).

\bibitem{Plis2016}
Plis, S.~M., Sarwate, A.~D., Wood, D., Dieringer, C., Landis, D., Reed, C., Panta, S.~R., Turner, J.~A., Shoemaker, J.~M., Carter, K.~W., Thompson, P., Hutchison, K., and Calhoun, V.~D., ``{COINSTAC}: A privacy enabled model and prototype for leveraging and processing decentralized brain imaging data,'' {\em Frontiers in Neuroscience}~{\bf 10} (Aug. 2016).

\bibitem{Lyon2017-rd}
Lyon, L., ``Dead salmon and voodoo correlations: should we be sceptical about functional {MRI}?,'' {\em Brain}~{\bf 140},  e53 (Aug. 2017).

\bibitem{Bennett2010-gy}
Bennett, C.~M. and Miller, M.~B., ``How reliable are the results from functional magnetic resonance imaging?,'' {\em Ann. N. Y. Acad. Sci.}~{\bf 1191},  133--155 (Mar. 2010).

\bibitem{SZUCS2020117164}
Szucs, D. and Ioannidis, J.~P., ``Sample size evolution in neuroimaging research: An evaluation of highly-cited studies (1990–2012) and of latest practices (2017–2018) in high-impact journals,'' {\em NeuroImage}~{\bf 221},  117164 (2020).

\bibitem{Turner2018-vq}
Turner, B.~O., Paul, E.~J., Miller, M.~B., and Barbey, A.~K., ``Small sample sizes reduce the replicability of task-based {fMRI} studies,'' {\em Commun. Biol.}~{\bf 1},  62 (June 2018).

\bibitem{Sayegh2014QualityOE}
Sayegh, P., Arentoft, A., Thaler, N.~S., Dean, A.~C., and Thames, A.~D., ``Quality of education predicts performance on the wide range achievement test-4th edition word reading subtest.,'' {\em Archives of clinical neuropsychology : the official journal of the National Academy of Neuropsychologists}~{\bf 29 8},  731--6 (2014).

\bibitem{Pervaiz2020-fm}
Pervaiz, U., Vidaurre, D., Woolrich, M.~W., and Smith, S.~M., ``Optimising network modelling methods for {fMRI},'' {\em Neuroimage}~{\bf 211},  116604 (May 2020).

\bibitem{Gichoya2022-kt}
Gichoya, J.~W., Banerjee, I., Bhimireddy, A.~R., Burns, J.~L., Celi, L.~A., Chen, L.-C., Correa, R., Dullerud, N., Ghassemi, M., Huang, S.-C., Kuo, P.-C., Lungren, M.~P., Palmer, L.~J., Price, B.~J., Purkayastha, S., Pyrros, A.~T., Oakden-Rayner, L., Okechukwu, C., Seyyed-Kalantari, L., Trivedi, H., Wang, R., Zaiman, Z., and Zhang, H., ``{AI} recognition of patient race in medical imaging: a modelling study,'' {\em Lancet Digit. Health}~{\bf 4},  e406--e414 (June 2022).

\bibitem{Li2022-rd}
Li, J., Bzdok, D., Chen, J., Tam, A., Ooi, L. Q.~R., Holmes, A.~J., Ge, T., Patil, K.~R., Jabbi, M., Eickhoff, S.~B., Yeo, B. T.~T., and Genon, S., ``Cross-ethnicity/race generalization failure of behavioral prediction from resting-state functional connectivity,'' {\em Sci. Adv.}~{\bf 8},  eabj1812 (Mar. 2022).

\bibitem{Wang2021-zs}
Wang, Z., Xin, J., Wang, Z., Yao, Y., Zhao, Y., and Qian, W., ``Brain functional network modeling and analysis based on {fMRI}: a systematic review,'' {\em Cogn. Neurodyn.}~{\bf 15},  389--403 (June 2021).

\bibitem{ds004144:1.0.2}
Balducci, T., Rasgado-Toledo, J., Valencia, A., van Tol, M.-J., Aleman, A., and Garza-Villarreal, E.~A., ``"a behavioral, clinical and brain imaging dataset with focus on emotion regulation of females with fibromyalgia",'' (2022).

\bibitem{Markiewicz2021-wj}
Markiewicz, C.~J., Gorgolewski, K.~J., Feingold, F., Blair, R., Halchenko, Y.~O., Miller, E., Hardcastle, N., Wexler, J., Esteban, O., Goncavles, M., Jwa, A., and Poldrack, R., ``The {OpenNeuro} resource for sharing of neuroscience data,'' {\em Elife}~{\bf 10} (Oct. 2021).

\bibitem{ds004775:1.1.1}
Weber, R., Hopp, F.~R., Eden, A., and Lee, K., ``"vicarious punishment of moral violations in naturalistic drama narratives predicts cortical synchronization",'' (2023).

\bibitem{Glessner2010-xg}
Glessner, J.~T., Reilly, M.~P., Kim, C.~E., Takahashi, N., Albano, A., Hou, C., Bradfield, J.~P., Zhang, H., Sleiman, P. M.~A., Flory, J.~H., Imielinski, M., Frackelton, E.~C., Chiavacci, R., Thomas, K.~A., Garris, M., Otieno, F.~G., Davidson, M., Weiser, M., Reichenberg, A., Davis, K.~L., Friedman, J.~I., Cappola, T.~P., Margulies, K.~B., Rader, D.~J., Grant, S. F.~A., Buxbaum, J.~D., Gur, R.~E., and Hakonarson, H., ``Strong synaptic transmission impact by copy number variations in schizophrenia,'' {\em Proc. Natl. Acad. Sci. U. S. A.}~{\bf 107},  10584--10589 (June 2010).

\bibitem{Satterthwaite2014NeuroimagingOT}
Satterthwaite, T.~D., Elliott, M.~A., Ruparel, K., Loughead, J., Prabhakaran, K., Calkins, M.~E., Hopson, R., Jackson, C., Keefe, J., Riley, M., Mentch, F.~D., Sleiman, P. M.~A., Verma, R., Davatzikos, C., Hakonarson, H., Gur, R.~C., and Gur, R.~E., ``Neuroimaging of the philadelphia neurodevelopmental cohort,'' {\em NeuroImage}~{\bf 86},  544--553 (2014).

\bibitem{Power2011FunctionalNO}
Power, J.~D., Cohen, A.~L., Nelson, S.~M., Wig, G.~S., Barnes, K.~A., Church, J.~A., Vogel, A.~C., Laumann, T.~O., Miezin, F.~M., Schlaggar, B.~L., and Petersen, S.~E., ``Functional network organization of the human brain,'' {\em Neuron}~{\bf 72},  665--678 (2011).

\bibitem{Tamminga2014-pp}
Tamminga, C.~A., Pearlson, G., Keshavan, M., Sweeney, J., Clementz, B., and Thaker, G., ``Bipolar and schizophrenia network for intermediate phenotypes: Outcomes across the psychosis continuum,'' {\em Schizophr. Bull.}~{\bf 40},  S131--S137 (Mar. 2014).

\bibitem{8857902}
Abrol, A., Rokham, H., and Calhoun, V.~D., ``Diagnostic and prognostic classification of brain disorders using residual learning on structural mri data,'' in [{\em 2019 41st Annual International Conference of the IEEE Engineering in Medicine and Biology Society (EMBC)}{\nolinebreak\hspace{0.1em}]},   4084--4088 (2019).

\bibitem{Chen2018-py}
Chen, Q., Kong, Y., Gao, W., and Mo, L., ``Effects of socioeconomic status, parent--child relationship, and learning motivation on reading ability,'' {\em Front. Psychol.}~{\bf 9} (July 2018).

\end{thebibliography}
\bibliographystyle{spiebib} 

\end{document}